\documentstyle[12pt]{article}
\begin{document}
\hsize=15.5cm
\textheight=24.5cm
\addtolength{\topmargin}{-90pt}
\large

\begin{center}

{ \bf GAUGE FIELDS WITH QUAZINILPOTENT GAUGE GROUP.} \\
     Arthur M. Aslanyan \\
     Kazan State University, Russia
\end{center}

\bigskip
\noindent
 { \bf Abstract}

     We investigate  non-linear  generalization  of  Maxwell  theory
of
el\-ec\-tro\-mag\-ne\-tic field
keeping the gauge invariance of Lagrangian.  New
theory,  which is standard Yang-Mills theory,  is based on
Harmonic Oscillator $HO(N,R)$ gauge group.  It's a solvable  Lie
group
with nilpotent normal subgroup of codimension 1. We wright down
the Yang-Mills equation and point out their pecularities and
connection
with standard Maxwell theory.

\bigskip
\noindent
{ \bf Field equations and interpretation.  }

 We  shall  operate  with
so-called Harmonic Oscillator Lie algebra $ho(4,R)$, which
is quazinilpotent and   admitts   non-degenerate   invariant
bilinear form:\\
$ b:L\times L \rightarrow R $,
such that $\forall x,y,z \in L$ the following condition:\\
$$ b(ad_z(x),y)+b(x,ad_z(y))=0    $$
is hold true.

  Consider a phase space of harmonic oscilator with generalized
variables:$p,q$ and Hamiltonian:$H=\frac{1}{2} (p^2+q^2)$. It's known
that a set $<1,p,q>$ forms a Lie algebra in respect to Poisson
brackets:
$$
\cases{
        [1,p]=[1,q]=0, & $ $ \cr
        [p,q]=1, & $ $}
$$
which is a nilpotent Lie algebra called Heisenberg algebra.

   If we join them a Hamiltonian the new set $<1,p,q,H>$ will form
new algebra:
$$
\cases{
       [1,p]=[1,q]=0, & $ $ \cr
       [p,q]=1,  & $ $ \cr
       [H,p]=q,  & $ $ \cr
       [H,q]=-p,  & $ $ }
$$
which is  a quazinilpotent Lie algebra and contains Heisenberg algebra
as an ideal. Indeed it's a semi-simple sum of 3-dimensional
Heisenberg algebra $<1,p,q>$ and one-dimensional algebra $<H>$.
This algebra was generalized in papers~\cite{gav}(See
also~\cite{diplom}).
The main property of this algebra is that it admitts non-degenerate
invariant bilinear form:
$$ <u,v>= u^sv^s + u^1v^4 + u^4v^1 + zu^4v^4  $$
Here $s,r,t,  ...=\overline{(2,3)}$, $z-$ any real constant
and $u^1,u^s,u^4-$ are the
components of the element u of Lie algebra in the basis $<1,p,q,H>$:
$$ u=u^1 1+u^2 p+u^3 q+u^4 H.    $$
Due to the existence of this form we can construct
for the gauge fields $\hat A_\alpha$ a Lagrangian
$L=\frac{1}{4}<\hat F_{\alpha \beta},  \hat  F^{\alpha \beta}>$
which  extremals  are the Yang-Mills equations exactly!\\
  Here hat  $\hat A_\alpha$ means that it belongs to the matrix
representation of gauge Lie algebra,  and $\alpha,\beta,\gamma,  ...-$
are the indices on  the
bundle (Minkowski) manifold.\\
$ \hat F_{\alpha \beta}=\partial_\alpha \hat A_\beta -
\partial_\beta \hat A_\alpha
      + [\hat A_\alpha, \hat A_\beta] -$ curvature tensor of gauge
field
     $\hat A_\alpha$,\\

Yang-Mills equations are:
$$
\cases{
 \partial^{\alpha} F^1_{\alpha \beta}+\omega_{st}A^{s
\alpha}F^t_{\alpha
\beta}=0, & $ $ \cr
 \partial^{\alpha}F^s_{\alpha \beta}+2\omega_{st}A^{\alpha
[t}F^{4]}_{\alpha
\beta}=0, & $ $ \cr
 \partial^{\alpha}F^4_{\alpha \beta}=0, & $ $ }
$$
coupled with system
$$
\cases{
  F^1_{\alpha \beta}=2\partial_{[ \alpha}A^1_{\beta ]}+\omega_{st}
        A^s_{\alpha}A^t_{\beta}, & $ $ \cr
  F^s_{\alpha \beta}=2\partial_{[ \alpha}A^s_{\beta ]}+2\omega_{st}
        A^4_{\alpha}A^t_{\beta}, & $ $ \cr
  F^4_{\alpha \beta}=2\partial_{[ \alpha}A^4_{\beta ]}. & $ $ }
$$
Here $\omega_{st}=-\omega_{ts}.$
  The system is semi-splitted in three parts.
  We see that $A^4$ component is a pure Maxwell field.
  We substitute this field to the second part and find the $A^s$
 components.The first part can be rewrighted as  a  Maxwell  equations
 with sources:
$$
\cases{
 \varphi_{\alpha \beta}=2\partial_{[ \alpha} A^1_{ \beta ]} , & $ $
\cr
     \partial^{\alpha}\varphi_{\alpha \beta}=J_{\beta}. & $ $ }
$$
 where $J_{\beta}=-
\omega_{st}\partial^{\alpha}(A^s_{\alpha}A^t_{\beta})
-\omega_{st}A^{s \alpha}F^t_{\alpha \beta}-$ "sources".

  Note that center of $ho(N,R)$ is $<1>$ and from physical point of
view it has no useful information, because the element ${1}$ was
coupled
to the  set ${p,q}$ to complete it to the Lie algebra,  what is
simply a
mathematical trick. It brings to the idea that component $A^1$
don't represent the real physical field, and putting it to be trivial
we can regard the first equation like a constrain on sources, more
precisely
on physical fields $A^2, A^3$.

     To wright down the gauge transformations for this theory
we have first to construct a Lie group with $ho(4,R)$ Lie algebra.
We call this group $HO(4,R)$ a harmonic oscilator Lie
group~\cite{aslgav}.
There are two types  of  gauge  transformations  (connected  with
semi-simple splitting of algebra into two subalgebras):\\
 First type:
$$
\cases{
    A^1 \rightarrow A^1, & $ $ \cr
    A^s \rightarrow (e^{-\lambda^4 \omega})^s_t A^t, & $ $ \cr
    A^4 \rightarrow A^4+\partial \lambda^4. & $ $}
$$
   Transformation for $A^4$ is usual gauge transformation for
Maxwell field (generally for any abelian gauge field). And
transformation for $A^s$ are simply rotation in the ${p,q}$
plane at angle $\lambda^4$:
$$
\cases{
  A^2 \rightarrow \,\,\cos(\lambda^4)A^2+\sin(\lambda^4)A^3, & $ $ \cr
  A^3 \rightarrow -\sin(\lambda^4)A^2+\cos(\lambda^4)A^3. & $ $ }
$$

  Second type:
$$
\cases{
    A^1 \rightarrow A^1+ \partial \lambda^1 + \omega_{st}\partial
 \lambda^s \lambda^t +(\lambda^s \lambda^s)A^4, & $ $ \cr
  A^s \rightarrow A^s+\partial \lambda^s+\omega_{st}\lambda^t A^4,
& $ $ \cr
  A^4 \rightarrow A^4. & $ $}
$$

 \bigskip
 \noindent
 {\bf Conclusion. }
 We see that putting $p,q-$ components of gauge field to zero
we come to standard Maxwell theory for $A^4-$ component. Then there
are two aspects of this theory.

The local is connected with
the physical meaning of $p,q-$ components which satisfy linear
equations and  non-linear constraints.  These components are not
independent because of the gauge transformations and we can well
 put one  of  them  (say $A^3$ ) to zero,  providing the gauge for the
usual Maxwell field $A^4$ being fixed. In this  case  the  constraints
will be  satisfied  automatically  and we come to the ordinary
two-component Maxwell theory with linear constraints on one of
the component.

  Another aspect is global. We know that in the Maxwell theory
all regular in $R^3$ monopole solutions are trivial. It turns out that
the same is  hold true for $HO(4,R)$-theory, but it's much more
difficult
to prove this~\cite{asl}. In Maxwell theory we are forced to regard
singular monopole  solutions (like Dirac magnetic monopole),  but it's
energy is infinite.  Does it valid for the $HO(4,R)-$ theory?  We
just
can claim the same thing in the case when Maxwell sector is trivial
($A^4=0$)~\cite{asl}, but it's not clear in the general case.

    And the last question: is it possible for non-trivial
$A^s, A^4$ field to give a zero energy? In this case we can speak
about
energyless electrodynamics, which can lead to the interesting physical
effects.
to

\end{document}